\documentclass{elsart}
\usepackage{graphics}
\usepackage{psfrag}
\usepackage{epsfig}
\usepackage{epsf}
\usepackage{float}
\usepackage{amssymb,stmaryrd,latexsym}
\usepackage{amscd}
\usepackage{mathrsfs}
\usepackage{color}

\newcommand{\be}{\begin{eqnarray}}
\newcommand{\ee}{\end{eqnarray}}
\newcommand{\beq}{\begin{equation}}
\newcommand{\eeq}{\end{equation}}

\definecolor{grey}{rgb}{0.3,0.3,0.3}

\usepackage{amsmath}
\usepackage{amssymb}
\usepackage{epsfig}

\begin{document}
\hfill{\small FZJ--IKP(TH)--2010--14, HISKP-TH-10-13}

\begin{frontmatter}
\title{Two- and three-body structure of the $Y(4660)$}

\author{P.~Hagen$^{1}$, H.-W.~Hammer$^1$, and C.~Hanhart$^{2}$}

{\small $^1$ Helmholtz-Institut f\"{u}r Strahlen- und Kernphysik (Theorie) and 
} \\ 
{\small Bethe Center for Theoretical Physics, Universit\"at Bonn, D-53115 Bonn, Germany}\\
{\small $^2$ Institut f\"{u}r Kernphysik,  (Theorie), Institute for Advanced
Simulation}\\
{\small and J\"ulich Center for Hadron Physics,}\\
{\small Forschungszentrum J\"ulich,  D--52425 J\"{u}lich, Germany} \\

\begin{abstract}
\noindent 
We study general features of three-meson bound states using the 
$Y(4660)$ as an example.
Here the $Y(4660)$ is assumed to be either a two-body bound 
state of the $f_0(980)$, itself a bound state of $K$ and $\bar{K}$,
and the $\psi'=\psi(2s)$, or a three-body bound state
of $\psi'$, $K$, and $\bar{K}$.
In particular, we investigate in detail the interplay of the various 
scales inherent in the problem, namely the $f_0$ binding energy, 
the $Y$ binding energy, and the $K\psi'$ scattering length. 
This allows us to understand under which circumstances the substructure of 
the $f_0(980)$ can be neglected in the description of the $Y(4660)$.
\end{abstract}

\end{frontmatter}

{\bf 1. Introduction:} In recent years a large number of new charmonium states
has been discovered experimentally, most of them showing properties in vast
conflict with the predictions of the quark model~\cite{reviews}.  The
structure of these states presents a serious challenge for our understanding
of QCD.  Since many of these new states are located near thresholds, it
appeared natural to propose a molecular nature for those. The prime example is
the $X(3872)$ located right at the $\bar D^0 D^{*0}$ threshold, proposed to be
a molecule composed of these mesons~\cite{Xmol} or at least a virtual
state~\cite{ourX}. However, this picture is not uniformly accepted. It was,
for example, challenged on the basis of production data from $\bar
pp$~\cite{antonio}. A rebuttal to this challenge was given in
Ref.~\cite{ericbarpp}.  In the light quark sector, the light scalar mesons
located below 1 GeV are also proposed to be molecules, namely the $f_0(600)$
--- the $\sigma$ meson --- the $f_0(980)$, $a_0(980)$, and
$\kappa(900)$~\cite{lightscalarmols,lightscalarcritcrit}. Because of the
relatively large widths involved, especially for the $\sigma$, and the larger
distance to the scattering thresholds, however, the situation is less clear in
this case. For a different viewpoint on the structure of these states,
see~\cite{lightscalarcrit1,lightscalarcrit2}.

These examples illustrate the importance of a model-independent method to determine the structure of such states close to 
scattering thresholds.
For two--body states located near a threshold there is a powerful method originally proposed by Weinberg for
the deuteron~\cite{wein}. This method was extended in Ref.~\cite{evidence} and
allows one to quantify the two--body molecular component.
A complementary formulation of this problem is given by effective range theory and
the effective field theory (EFT) for large scattering length. The latter framework can also be extended to
three- and four-body molecules \cite{ERE-EFT,HammerBraaten:2004}. Both methods effectively analyze the 
low-energy pole structure of the scattering amplitude in a model-independent way.

In this paper, we want to investigate to which extent it is possible to distinguish between two- and three-body molecules 
and which scales govern this distinction. In particular, we analyze the example of the 
$Y(4660)$ in detail. This state was proposed to be a $\psi'f_0(980)$ molecule based on an analysis with the Weinberg method~\cite{Guo:2008zg}
while the $f_0(980)$ itself was proposed to be a molecule of $\bar K$ and $K$ mesons~\cite{lightscalarmols,lightscalarcritcrit}. 
It is therefore important to elucidate whether the internal structure of the $f_0(980)$ can be neglected for the description
of the $Y(4660)$ or not. The answer to this question, of course, depends on the physical scales in this problem. For a 
deeply-bound $f_0(980)$, one expects that the internal structure can be neglected.
Performing numerically exact three-body calculations, we determine for which parameter 
ranges the $Y(4660)$ can be interpreted as a $\psi'f_0(980)$ molecule and for which ranges as a $\psi'K\bar K$ molecule.
We stress that we only investigate the interplay of a possible
two--body vs. three--body nature of the $Y(4660)$ in order to understand the
interplay of the various scales. No explicit compact component is
included for the $Y(4660)$ nor is the width of the $f_0(980)$ included. Thus, this work
is mainly of theoretical interest and we are not yet in the position to
compare with experimental line shapes.
We start with a brief description of our formalism and some analytical considerations
before we present and discuss the results of our three--body calculation. The paper ends with an outlook.

\begin{figure}
 \centering
 \psfrag{i}{$i$}
 \psfrag{j}{$j$}
 \psfrag{k}{$k$}
 \begin{tabular}{clcl}
  \vspace{4mm}
   & \raisebox{-0.375\height}{\includegraphics[scale=0.77]{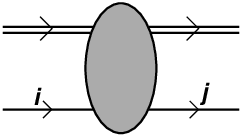}} & $=$ & \raisebox{-0.4\height}{\includegraphics[scale=0.77]{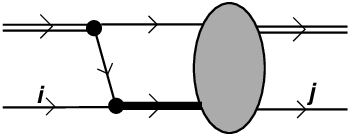}}\\
   \vspace{2mm}
   $+$ & \raisebox{-0.4\height}{\includegraphics[scale=0.77]{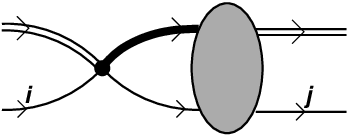}} & $+$ & \raisebox{-0.4\height}{\includegraphics[scale=0.77]{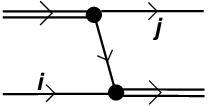}}
$\quad+\quad$\raisebox{-0.4\height}{\includegraphics[scale=0.77]{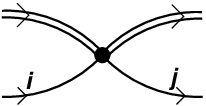}}\\
 \end{tabular}
 \caption{Diagrammatic representation of the integral equations used. The bare and full dimer fields are indicated by the double and thick lines, respectively. 
The equation for the latter is illustrated in Fig.~\ref{Graphic:Dimerprop}.}
 \label{Graphic:Amplitude}
\end{figure}

\begin{figure}
 \centering
 \psfrag{i}{$i$}
 \psfrag{j}{$j$}
 \psfrag{k}{$k$}
 \begin{tabular}{ccccc}
  \raisebox{-0.15\height}{\includegraphics[scale=0.9]{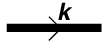}} & $=$ & \raisebox{-0.15\height}{\includegraphics[scale=0.9]{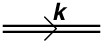}} 
& $+$ & \raisebox{-0.415\height}{\includegraphics[scale=0.9]{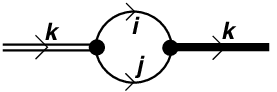}}\\
 \end{tabular}
 \caption{Diagrammatic representation for the full dimer propagator in channel $k$.}
 \label{Graphic:Dimerprop}
\end{figure}

{\bf 2. Formalism:} In this section, we briefly outline the approach used to
derive the amplitude for $\psi'f_0(980)$--scattering. For this purpose, we set
up a non-relativistic effective field theory (EFT) for three distinguishable
particles with different masses.\footnote{ A detailed description of the EFT
  for three identical bosons is given in~\cite{HammerBraaten:2004}.  The
  general case of distinguishable particles with different masses can be found
  in Ref.~\cite{HagenDip}.}  The underlying Lagrangian reads~\cite{HagenDip}
\begin{equation}
 \begin{split}
  \mathcal L &= \sum_{k=0}^2 \psi_k^\dagger \left( i \partial_t + \frac{\vec \nabla^2}{2m_k} \right) \psi_k - \sum_{k=0}^2 g_{k} 
\left( d_k^\dagger d_k - d_k^\dagger \psi_i \psi_j - d_k \psi_i^\dagger \psi_j^\dagger \right) \\
  &- h \sum_{k=0}^2 c_k d_k^\dagger d_k \psi_k^\dagger \psi_k\quad,
 \end{split}
 \label{Equation:Lagrangian}
\end{equation}
where the scalar functions $\psi_k$ and $d_k$ are particle- and dimer fields, respectively, and $m_k$ is the mass 
of particle $k$. The indices $i, j, k$ in Eq.~\eqref{Equation:Lagrangian} are always different from each other such 
that the mass of particle $k$ can also be labeled by $m_{ij}=m_{ji}$. This convention will be used below. Moreover,
a two-body state of particles $i$ and $j$ can be labeled either by the indices $ij$ or by the index $k$ ($i\neq j\neq k$).
The dimer fields are introduced for convenience and are not dynamical. 
An equivalent theory without dimer fields can be obtained by inserting the
classical equation of motion for the non--dynamical dimer fields.
Furthermore, $g_k$ and $h$ are coupling--constants, and the factors $c_k$ sum up to one.  
From the Lagrangian \eqref{Equation:Lagrangian}, we can 
directly derive the Feynman--rules of our theory in momentum space and construct the general amplitude 
$A_{ij}$ for particle--dimer scattering in the channel $i\to j$. The resulting equations are illustrated in Figs.~\ref{Graphic:Amplitude} 
and~\ref{Graphic:Dimerprop}.

After renormalizing the amplitude and performing a partial wave decomposition,
the inhomogeneous integral equation for the $l$--th partial wave reads
\begin{equation}
   \left[A_{ij}\right]_l\left(p,k\right) = \sum_{k=0}^2 \int_0^\Lambda \text{d}q \; \left[T_{ik}\right]_l\left(p,q\right)\;S_k\left(q \right) 
\left[A_{kj}\right]_l\left(q,k\right) \;-\; \left[T_{ij}\right]_l\left(p,k\right)\quad,
 \label{Equation:Amplitude}
\end{equation}
where $p$ and $k$ is the incoming and outgoing momentum, respectively. The kernel factorizes into the driving term
\begin{equation}
  \left[T_{ik}\right]_l\left(p,q\right) := (1-\delta_{ik})\frac{2\pi\;m_{ik}}{\mu_i\mu_k\sqrt{|a_i||a_k|}}\;\frac{1}{pq} Q_l\left(c_{ik}(p,q)\right) \;+\; 
\delta_{ik}\;\delta_{l0}\;\frac{H(\Lambda)}{\Lambda^2}
 \label{Equation:T}
\end{equation}
and the full dimer propagator
\begin{equation}
 S_k\left(q\right) := \frac{\mu_k|a_k|q^2}{2\pi^2}\left[\frac{1}{a_k} - \sqrt{2\mu_k\left(\frac{q^2}{2\tilde\mu_k}-E\right)-i\epsilon}\right]^{-1} \quad,
 \label{Equation:S}
\end{equation}
where 
\begin{equation}
\mu_k := \frac{m_i m_j}{m_i+m_j}\quad\mbox{ and }\quad
\tilde{\mu}_k:=\frac{(m_i+m_j) m_k}{(m_i+m_j)+m_k}
\label{redmass}
\end{equation}
are the reduced masses of the two-particle system labeled by $ij$ or by $k$ ($i\neq j\neq k$)
and the particle-dimer system labeled by $k$, respectively.
The Legendre functions of the second kind $Q_l$ and their arguments $c_{ik}(p,q)$ are defined by
\begin{equation}
 Q_l(c) \ := \ \frac{1}{2}\int_{-1}^1 dx \;
 \frac{P_l(x)}{c-x}\quad\text{and}\quad c_{ik}(p,q) \ :=
 \ \frac{m_{ik}}{pq}\left(E-\frac{p^2}{2\mu_k} - \frac{q^2}{2\mu_i} +
 i\epsilon\right) \ ,
 \label{Equation:Qc}
\end{equation}
where $P_l$ denotes the $l$-th Legendre polynomial.  We use standard solution techniques in order to solve the integral 
Eq.~\eqref{Equation:Amplitude} numerically. The apriori undetermined, dimensionless function $H(\Lambda)$, that by 
construction contributes only in the $s$--wave, is directly related to the parameter $h$ in Eq.~\eqref{Equation:Lagrangian}. 
The function $H(\Lambda)$ plays the role of a running coupling constant and is determined 
by the renormalization procedure up to an unknown three-body parameter~\cite{Bedaque:1998kg}.
We fix its value by imposing that a three--particle bound--state exists at the binding 
energy $E=-B$. This requires that the amplitude $A_{ij}$ in Eq.~\eqref{Equation:Amplitude} has a pole at this energy. 
However, any low-energy three-body observable can be chosen for this purpose. After fixing $H$,  
we are able to calculate the full particle--dimer 
scattering amplitude and therefrom, observables such as scattering lengths, cross sections and so forth can be predicted.

This theory represents the leading order of an EFT around the limit of large pair scattering 
lengths $a_k$~\cite{ERE-EFT,HammerBraaten:2004}.
Corrections are suppressed by powers of $r_0/a_k$ and $pr_0$, where $r_0$ is the range of the interaction
and $p$ is a typical momentum. If the scattering lengths are not sufficiently large, the theory can still be taken 
as a particular three-body model for hadronic molecules. As such, it can be used to test the limits of applicability of
the Weinberg approach to systems like the $Y(4660)$. By calculating the scattering length and the effective 
range, we are able to investigate the question, whether the three--particle bound--state can effectively be treated as a 
two--particle system, consisting of one particle and an elementary dimer. If the substructure of the dimer becomes relevant,
the full three--particle picture is necessary. Transferred to our specific example $Y(4660)$, this means that we want to 
distinguish between the two alternatives $\psi'f_0(980)$ and $\psi'K\bar K$. In the 
Weinberg formalism~\cite{wein}, the quantity $Z\in[0,1]$ is defined as
\begin{equation}
 Z \ := \  1-\int \text{d}p \ \Big| \, \langle \, \psi'f_0(p) \, | \, Y \, \rangle \,
 \Big|^2\quad \ ,
 \label{Equation:Zdef}
\end{equation}
where $Y$ denotes the wave function of the physical state. Thus, $Z$ measures the "non--two--particle"--fraction of 
the wave function. From the definition~\eqref{Equation:Zdef}, we directly conclude that  $Z\to0$ implies that the
$Y(4660)$ is effectively a two--body system with an elementary $f_0(980)$--dimer, whereas $Z\to1$ means that it has 
to be seen as something else,  in our case a three--body molecule. 
This method of distinction between two-- and three--particle molecules can of 
course also be applied to other candidates for hadronic molecules in order to reveal their few--particle nature.
The quantity $Z$ is directly related to the residue $Z_{pole}$ of the bound state pole in the two-particle 
scattering amplitude by $Z=1-Z_{pole}$. 

Performing a straightforward calculation within the Weinberg formalism, using the Lippmann--Schwinger equation and 
the effective range expansion, the quantity $Z$ can be related to the effective range parameters in two different 
ways~\cite{wein}. The two corresponding $Z$--factors, which we denote by
\begin{equation}
 \label{Equation:Zar}
 Z_a \ := \ \frac{a^{2\text{P}}_{\psi'f_0} - a_{\psi'f_0}}{a^{2\text{P}}_{\psi'f_0} - \frac{a_{\psi'f_0}}{2}} \qquad \text{and}\qquad  
Z_r \ := \ \frac{r_{\psi'f_0}}{r_{\psi'f_0}-a^{2\text{P}}_{\psi'f_0}}\quad,
\end{equation}
should be (approximately) equal if the Weinberg formalism is applicable. The equations (\ref{Equation:Zar}, \ref{Equation:a2T})
receive corrections from the finite range of the interaction. They are exact in the limit of vanishing binding energy of the $Y(4660)$ 
relative to the $\psi'f_0$-threshold with $Z=Z_a=Z_r$ kept fixed~\cite{wein}.  
The quantities $a_{\psi'f_0}$ and $r_{\psi'f_0}$ are the scattering length and the
effective range for $\psi'f_0$-scattering, respectively. Moreover, 
\begin{equation}
 a^{2\text{P}}_{\psi'f_0}=\frac{1}{\sqrt{2\ \mu_{\psi'f_0}B_{\psi'f_0}}}=\frac{1}{\sqrt{2\ \mu_{\psi'f_0}\left(B_{Y} - B_{f_0}\right)}}
 \label{Equation:a2T}
\end{equation}
is the scattering length within a pure two--particle picture. 

The positive binding energies are defined via
\begin{align}
 \label{Equation:BY}
 B_Y &:= m_{\psi'}+m_K+m_{\bar{K}}-m_Y\\
 \label{Equation:BPsif0}
 B_{\psi'f_0} &:= m_{\psi'}+m_{f_0}-m_Y\\
 B_{f_0} &:= m_K+m_{\bar{K}}-m_{f_0}\quad.
 \label{Equation:Bf0}
\end{align}
Thereby all quantities on the right sides of the equations in \eqref{Equation:Zar} can be calculated in the 
full three-body model.
Combining the two foregoing conditions $Z\in[0,1]$ and $Z=Z_a=Z_r$, we conclude that
\begin{equation}
 0\leq Z_a \approx Z_r \leq 1
 \label{Equation:consistency}
\end{equation}
should hold. Below we will use the validity of Eq.~\eqref{Equation:consistency} as a diagnostic tool 
to identify the range of applicability of the Weinberg method.

The full derivation of the equations~\eqref{Equation:Zar} can be found in~\cite{wein}. We omit it at this point, but outline 
its essential idea. Utilizing the binding energy between the particle $\psi'$ and $f_0(980)$ as defined in 
Eq.~\eqref{Equation:BPsif0}, the Schr\"{o}dinger equation can be applied in the form
\begin{equation}
 \hat H \, | \, Y \, \rangle= -B_{\psi'f_0} \, | \, Y \, \rangle \qquad\text{with} \qquad\hat H = \hat H_0+\hat V\quad.
 \label{Equation:H}
\end{equation}
Here $\hat V$ describes the interaction and $\hat H_0$ denotes the free--particle part of the Hamiltonian, such that
\begin{equation}
 \hat H_0 \, | \, \psi'f_0(p) \, \rangle = E(p) \, | \, \psi'f_0(p) \, \rangle\quad
 \label{Equation:H_0}
\end{equation}
holds, where $E(p)=p^2/(2\mu_{\psi'f_0})$ and $p$ is the relative momentum of the $\psi'$ and $f_0$ in the center-of-mass.
Employing the relations~\eqref{Equation:H} and~\eqref{Equation:H_0}, 
the momentum--integral appearing in the definition of $Z$~\eqref{Equation:Zdef} assumes the form
\begin{equation}
 \begin{split}
 \int \text{d}p \ \Bigl| \, \langle \,  \psi'f_0(p) \, | \, Y \, \rangle  \, \Bigr|^2 &=
 \int \text{d}p \ \Bigl| \, \langle \, \psi'f_0(p) \, | \, \frac{\hat H_0-\hat H}{E(p)+B_{\psi'f_0}} \, | 
\, Y \, \rangle \, \Bigr|^2\\
 &=\int \text{d}p \ \frac{|\langle\psi'f_0(p)|\hat V|Y\rangle|^2 }{(E(p)+B_{\psi'f_0})^2 }\quad.
 \end{split}
 \label{Equation:Z}
\end{equation}
In order to deduce the formulas~\eqref{Equation:Zar}, the crucial requirement of the Weinberg method is that
the numerator on the right-hand-side of Eq.~\eqref{Equation:Z} can be approximated by its value at zero momentum,
\begin{equation}
 \label{Equation:approx}
 \int \text{d}p \ \frac{ | \langle \psi'f_0(p) | \hat V | Y \rangle |^2 }{(E(p)+B_{\psi'f_0})^2 } \approx \int \text{d}p \ 
\frac{ | \langle   \psi'f_0(0) | \hat V | Y \rangle |^2 }{ (E(p)+B_{\psi'f_0})^2}\quad .
\end{equation}
For this to hold, the numerator has to vary slowly over the range of momenta contributing to the integral, i.e. 
\begin{equation}
 E(p)\,\lesssim\, B_{\psi'f_0} \qquad\Rightarrow\qquad p\,\lesssim\, \sqrt{2\mu_{\psi'f_0}B_{\psi'f_0}}\quad.
\end{equation}
In other words, the range of the form factor 
$\langle \psi'f_0(p) | \hat V | Y \rangle$ has to be much larger than the range of the 
denominator, $\sqrt{2\mu_{\psi'f_0}B_{\psi'f_0}}$.

\begin{figure}[tbp]
 \psfrag{i}{$i$}
 \psfrag{Y4660}{$Y(4660)$}
 \centering
 \includegraphics[scale=1,clip]{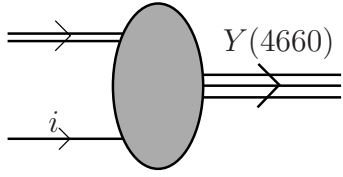}
 \caption{Feynman diagram for the form--factor $\langle d_i\,\psi_i(p) | \hat V | Y \rangle$ 
with a particle and an dimer of type $i$ in the incoming channel and $Y(4660)$ as the outgoing bound state.}
 \label{Graphic:FG_Formfactor_i}
\end{figure}

{\bf 3. Analytical Considerations:} 
Thus we have to understand the relevant momentum--scales of $\langle \psi'f_0(p) | \hat V | Y \rangle$. In the 
general case, with an incoming particle and dimer of type $i$ this form--factor 
$\langle \psi_i\, d_i(p)\, | \hat V | Y \rangle$ is proportional to the bound state amplitude $F_i$ as it is depicted in
Fig.~\ref{Graphic:FG_Formfactor_i}. It can be calculated by solving the homogeneous analog 
of Eq.~\eqref{Equation:Amplitude}, 
\begin{equation}
 \left[F_i\right]_0\left(p\right) = \sum_{k=0}^2 \int_0^\Lambda \text{d}q \, \left[T_{ik}\right]_0\left(p,q\right)\,
S_k\left(q \right) \left[F_k\right]_0\left(q\right)\quad,
 \label{Equation:Formfactor}
\end{equation}
where we have set $l=0$ for an $s$--wave state.
In the rest frame of the $Y(4660)$, the total energy is simply $E=-B_Y$. The quantity $p$ is the relative momentum 
between the incoming particles. Solving Eq.~\eqref{Equation:Formfactor} exactly requires the same effort as solving 
the scattering equation and can only be done numerically. Such numerical solutions will be presented below.

Our aim here is to investigate to what extent the underlying scales of 
the form factor $\langle \psi_i\, d_i(p)\, | \hat V | Y \rangle$
can be understood in a simple way. For this purpose, we approximate the full amplitude
$\left[F_k\right]_l\left(q\right)$ on the right-hand-side of  Eq.~\eqref{Equation:Formfactor} 
by a constant. In this point--like approximation, the $p$--dependence of 
$\langle \psi_i\, d_i(p)\, | \hat V | Y \rangle$ emerges from the integral
\begin{equation}
 \begin{split}
  &\int_0^\Lambda \text{d}q \, \left[T_{ik}\right]_0\left(p,q\right)\,S_k\left(q \right) \\ 
  =&\int_0^\Lambda \text{d}q \left\{(1-\delta_{ik})\frac{m_{ik}}{\pi\mu_i}\sqrt{\frac{|a_k|}{|a_i|}} \frac{q}{p} Q_0\left(c_{ik}(p,q)\right)+ \delta_{ik}\;
\frac{H(\Lambda)}{\Lambda^2}\frac{\mu_k|a_k|q^2}{2\pi^2}\right\}\\
  &\qquad\quad\times \left[\frac{1}{a_k} - \sqrt{2\mu_k\left(B_Y+\frac{q^2}{2\tilde\mu_k}\right)}\right]^{-1} \quad,
 \end{split}
 \label{Equation:TS}
\end{equation}
where we have inserted the definitions~\eqref{Equation:T} and~\eqref{Equation:S}. 
Using~\eqref{Equation:Qc}, we deduce that for $E=-B_Y$ the quantity $c:=c_{ik}(p,q)$ is always smaller than $-1$ and 
approaches 
$-\infty$ for $p\to0$. Hence a Taylor--expansion of $Q_0$ around $1/c=0$ can be performed and we approximate:
\begin{equation}
  Q_0(c)=\frac12\ln\frac{1+1/c}{1-1/c}
\approx\frac1c\quad.
\end{equation}
Inserting this result into~\eqref{Equation:TS}, the momentum dependence of $\langle \psi_i\, d_i(p)\, | 
\hat V | Y \rangle$ for $p\to0$ has the form:
\begin{equation}
 \begin{split}
 \propto \,&(1-\delta_{ik})\int_0^\Lambda \text{d}q \,\frac{q^2}{B_Y+\frac{p^2}{2\mu_k}+\frac{q^2}{2\mu_i}}\,
\frac{1}{\frac{1}{a_k} - \sqrt{2\mu_k\left(B_Y+\frac{q^2}{2\tilde\mu_k}\right)}}\quad.
 \end{split}
\end{equation}
This expression is nearly constant for momenta
\begin{equation}
 0\leq p\lesssim\sqrt{2\mu_k B_Y} \quad,
\end{equation}
so that the range of the form--factor can be estimated by $\sqrt{2\mu_k B_Y}$. Coming back to our specific situation 
with $\psi'$ and the $f_0$--dimer, the range of $\langle \psi'f_0(p) | \hat V | Y \rangle$ is estimated as
$\sqrt{2\mu_{\psi'K} B_Y}$. From this we are able to formulate the condition
\begin{equation}
 \sqrt{2\mu_{\psi'K} B_Y}\gg\sqrt{2\mu_{\psi'f_0} B_{\psi'f_0}}
\end{equation}
for the validity of Eq.~\eqref{Equation:approx}. Since the reduced mass terms $\mu_{\psi'K}$ and $\mu_{\psi'f_0}$ are 
both of the order $m_K$ and $B_{\psi'f_0}=B_Y-B_{f_0}$, this is equivalent to:
\begin{equation}
 B_Y\gg B_Y-B_{f_0}\quad\Leftrightarrow \quad 1\gg 1-b_0\quad,
 \label{Equation:bcond}
\end{equation}
where we have introduced the dimensionless parameter $b_0:=B_{f_0}/B_Y\in(0,1)$, which is the
binding energy of the $f_0$--dimer relative to the binding energy of the $Y(4660)$. 
Thus in this approximation the only relevant scale that controls the 
applicability of the Weinberg--method is $b_0$ which has to be close to $1$, meaning that $f_0$ has to be a deeply 
bound state or an elementary particle. 

We stress that $b_0$ can not be the only relevant scale in this problem. In particular, the scale characterizing
the $K\psi'$ and  $\bar K\psi'$ interactions does not enter at all in this simple picture. In the full solution of the three--body
problem where the $Y(4660)$ emerges as a dynamically generated three--particle--state, the situation will be
more complex. We will investigate this question below by solving Eq.~\eqref{Equation:Amplitude} numerically. 

{\bf 4. Results and discussion:\label{section3}} Using our formalism, we now present some results for 
$\psi'f_0(980)$--scattering within our three--body approach for the $Y(4660)$. In order to calculate observables, 
we first have to fix the 3 masses and scattering lengths, appearing in~\eqref{Equation:T}, \eqref{Equation:S} and 
\eqref{Equation:Qc}. Ignoring the -- for our purpose -- insignificant widths and errors, we take PDG values for the 
3 masses by setting $m_{\psi'}=3686.1$~MeV and $m_K=m_{\bar{K}}=(m_{K^+}+m_{K^0})/2=495.6$~MeV. The corresponding 
two--particle scattering lengths are currently unknown, but for the kaon--kaon--interaction we can determine 
$a_{K\bar{K}}=1/\sqrt{2\mu_{K\bar{K}}B_{f_0}}$ by employing the analogue of Eq.~\eqref{Equation:a2T}.
This corresponds to treating the $f_0$ as shallow bound state of $K$ and $\bar{K}$. 
For $a_{\psi'K}$ and $a_{\psi'\bar{K}}$, we have neither experimental data nor predictions from lattice 
calculations. Due to the absence of a two--particle bound--state and owing to symmetry reasons we can, however, 
conclude that $a_{\psi'K}=a_{\psi'\bar{K}}<0$ should hold. Another restriction on this quantity's magnitude comes 
from the fact that in $\psi'K$--scattering no quark exchange is possible in contrast to $DK$--scattering, for instance. 
Therefore, its interaction should be suppressed compared to the one in the latter case, implying that $|a_{\psi'K}|$ 
should be smaller than typical, non--resonant scattering lengths for $DK$-processes. 
In~\cite{Hanhart:DK09}, $DK$ scattering lengths were calculated, using unitarised chiral perturbation theory 
and in Ref.~\cite{lattice} they were extracted from lattice QCD simulations. 
Both analyses agree and provide values of the order of $0.1$~fm. 
Thus we demand $-0.1\text{ fm} \lesssim a_{\psi'K}=a_{\psi'\bar{K}}<0$~fm. 

We note that, given this assumption,   
$a_{\psi'K}$ is not large compared to the range of forces such that the EFT for large scattering lengths is
strictly not applicable. In this study, however, we use the EFT as a specific three-body model
that can be solved numerically in order to test the range of applicability of the Weinberg method.
In relation to the 
corresponding thresholds in Eq.~\eqref{Equation:BY} and~\eqref{Equation:Bf0} the masses $m_Y=4664\pm11\pm5$~MeV 
and $m_{f_0}=980\pm10$~MeV have large errors respectively so that we are still allowed to vary the binding energies in 
a region $0<B_{f_0}<B_Y\lesssim20$~MeV and be consistent with experiment.

\begin{figure}[htbp]
 \centering
 \psfrag{b0}{$b_0$}
 \psfrag{aPsif0[fm]}{$a_{\psi'f_0}$ [fm]}
 \psfrag{BY=}{$B_Y=$}
 \psfrag{5 MeV 3TM}{\scriptsize 5 MeV 3P}
 \psfrag{5 MeV 2TM}{\scriptsize 5 MeV 2P}
 \psfrag{10 MeV 3TM}{\scriptsize 10 MeV 3P}
 \psfrag{10 MeV 2TM}{\scriptsize 10 MeV 2P}
 \psfrag{20 MeV 3TM}{\scriptsize 20 MeV 3P}
 \psfrag{20 MeV 2TM}{\scriptsize 20 MeV 2P}
 \includegraphics[scale=0.56,clip]{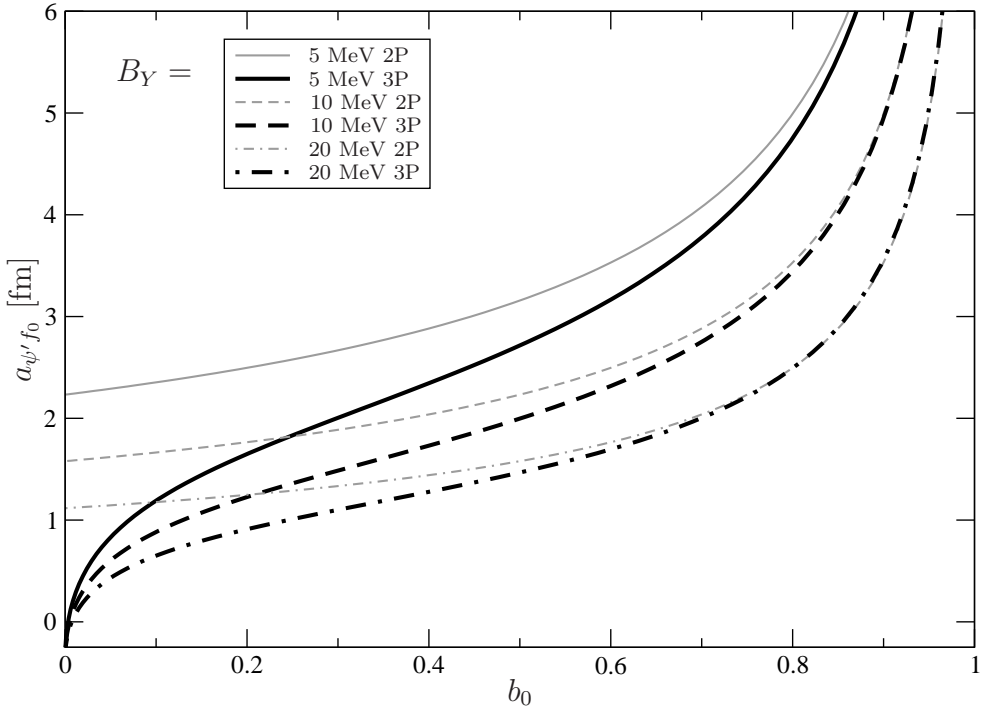}
 \caption{Scattering length for the $\psi'f_0(980)$--process as a function of the relative binding energy 
$b_0=B_{f_0}/B_Y$ for different values of $B_Y$, the binding energy of the $Y$
   with respect to the $\bar KK\psi'$ threshold (see Eq.~(\ref{Equation:BPsif0})). Our numerical values within the three--particle model (3P) are compared 
to the ones in a simple two--particle model (2P). The unknown scattering length
$a_{\psi'K}=a_{\psi'\bar{K}}$ is set to $-0.1$ fm.}
 \label{Graphic:aPsif0}
\end{figure}

\begin{figure}[htbp]
 \centering
 \psfrag{b0}{$b_0$}
 \psfrag{rPsif0[fm]}{$r_{\psi'f_0}$ [fm]}
 \psfrag{BY=}{$B_Y=$}
 \psfrag{5 MeV x}{\scriptsize 5 MeV}
 \psfrag{10 MeV}{\scriptsize 10 MeV}
 \psfrag{20 MeV}{\scriptsize 20 MeV}
 \centering
 \includegraphics[scale=0.56,clip]{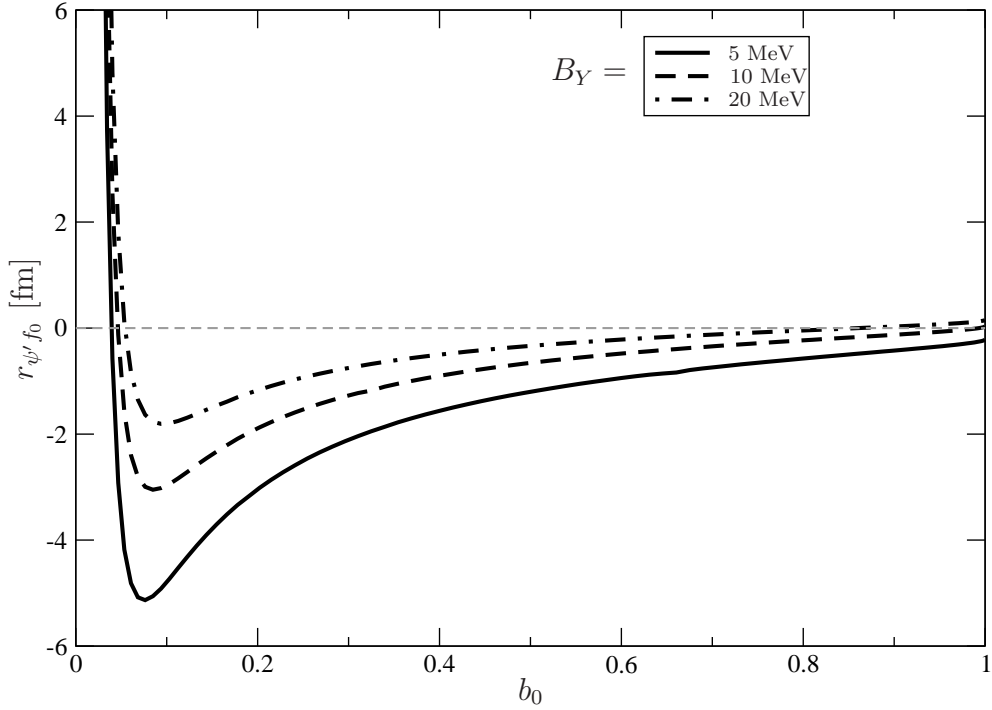}
 \caption{Effective range for $\psi'f_0(980)$--scattering as a function of the relative binding energy 
$b_0=B_{f_0}/B_Y$ for different values of $B_Y$. The unknown scattering length 
$a_{\psi'K}=a_{\psi'\bar{K}}$ is set to $-0.1$ fm.}
 \label{Graphic:rPsif0}
\end{figure}

\begin{figure}[htbp]
 \centering
 \psfrag{b0}{$b_0$}
 \psfrag{Za}{$Z_a$}
 \psfrag{b1=}{$b_1=$}
 \psfrag{1e+1}{\small $10^1$}
 \psfrag{1e+3}{\small $10^3$}
 \psfrag{1e+5}{\small $10^5$}
 \psfrag{1e+7}{\small $10^7$}
 \psfrag{b1}{$b_1$}
 \centering
 \includegraphics[scale=0.56,clip]{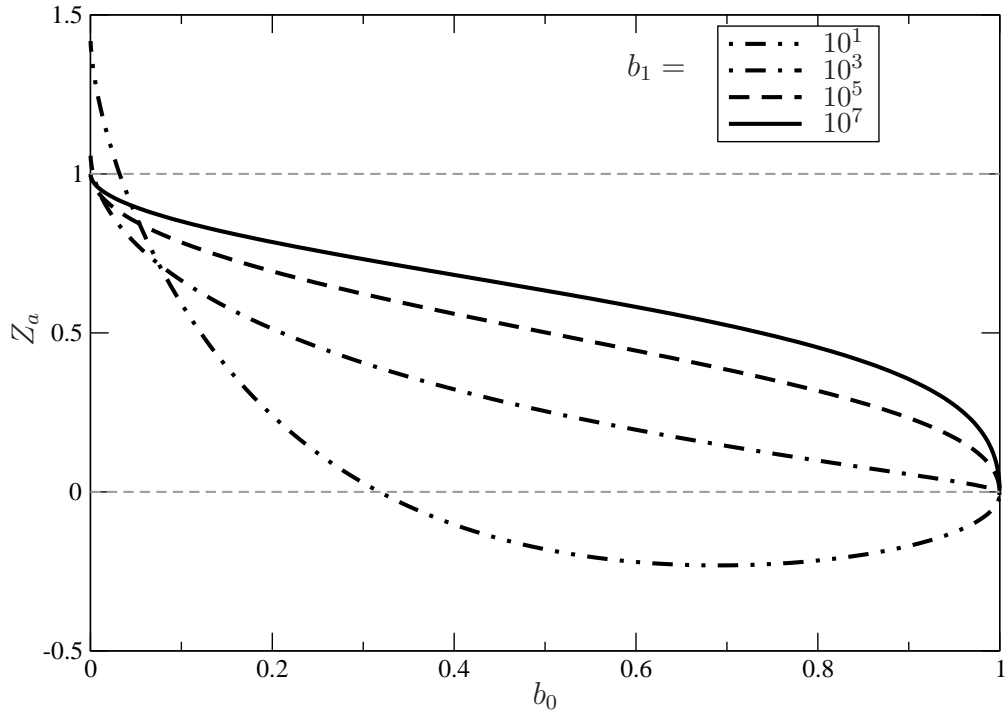}
 \caption{$Z_a$ from Eq.~\eqref{Equation:Zar} as a function of the relative binding energy $b_0=B_{f_0}/B_Y$ for different 
values of $b_1=B_1/B_Y$. 
}
 \label{Graphic:Z}
\end{figure}

In Fig.~\ref{Graphic:aPsif0}, the numerically calculated $\psi'f_0(980)$--scattering length within 
the three--particle model (3P) is depicted for several values of the total binding energy $B_Y$. It is also compared to 
the one of Eq.~\eqref{Equation:a2T} in a hypothetical two--particle system (2P) with no substructure in the 
$f_0(980)$--dimer. The variation in $B_{f_0}$ is expressed in a dimensionless relative binding energy $b_0$. The 
unknown quantity $a_{\psi'K}=a_{\psi'\bar{K}}$ is set to $-0.1$~fm.\footnote{For the numerical calculation, we typically use
$N=100-200$ mesh points and a cutoff $\Lambda=10^{11}$~MeV. The obtained results are independent of this
choice, however, as long as $\Lambda$ is large compared to all momentum scales in the problem.}
Our numerical results for the $\psi'f_0(980)$--scattering length 
confirm the behavior deduced from the analytical considerations above (cf.~Eq.~\eqref{Equation:bcond}).
For a weakly bound $f_0(980)$--dimer, that is $b_0\to0$, we see deviations between the two models, whereas for a 
strongly bound dimer with $b_0\to1$ the results nearly coincide. In the latter case $a_{\psi'f_0}$ diverges to $+\infty$. 
We also calculated the effective range in Fig.~\ref{Graphic:rPsif0} for the same parameters. It diverges at values where 
the scattering length vanishes. 
Unfortunately, neither experimental data nor lattice calculations are available at present and no comparison
to data for $\psi'f_0(980)$--scattering can be made.

By applying the Weinberg--formalism, described in the previous section, we are also able to investigate the question at 
which scales $Y(4660)$ can effectively be seen as a two--body molecule consisting of elementary particles $f_0(980)$ and 
$\psi'$ and where the $f_0(980)$--substructure has to be taken into account. 
Since the only undetermined scales in our 
system are $B_Y$, $B_{f_0}$ and $a_{\psi'K}$, we expect the dimensionless quantities $Z_a$ and $Z_r$, as 
defined in the equations~\eqref{Equation:Zar}, only to depend on 2 dimensionless ratios of these 3 parameters. 
We have already shown above that the relative binding energy of the $f_0$ and $Y(4660)$, $b_0$, plays an important role
in determining the structure of the $Y(4660)$. For the second ratio it is natural to choose
$b_1:=B_1/B_Y:=1/(2\mu_{\psi'K}\, a_{\psi'K}^2 \, B_Y)$.
Since the scattering length $a_{\psi'K}$ is negative, the energy scale $B_1$ characterizes a virtual state
in the $\psi'K$--channel but not a bound state. The functions $Z_a(b_0)$ for discrete values of $b_1$, varied 
over a wide range, are displayed in Fig.~\ref{Graphic:Z}. 
We find that $Z_a$ strongly depends on $b_0$ and $b_1$. For $b_0\to0$, $Z_a$ reaches the value of $1$ so that $Y(4660)$ has to be 
considered a three--body system. In the limit $b_0\to1$, $Z_a$ approaches $0$ as the $f_0(980)$--dimer gets maximally bound. 
In this case, the $Y(4660)$ can  be seen as a two--particle system consisting of the elementary 
particles $\psi'$ and $f_0(980)$.

\begin{figure}[htbp]
 \centering
 \psfrag{b0}{$b_0$}
 \psfrag{sqrtb1}{$\sqrt{b_1}$}
 \psfrag{Za<0}{$Z_a<0$}
 \psfrag{0<Za<1}{$0<Z_a<1$}
 \psfrag{Za>1}{$Z_a>1$}
 \psfrag{Za=Zr}{$Z_a\approx Z_r$}
 \psfrag{Za<>Zr}{$Z_a\neq Z_r$}
 \psfrag{b1<225}{\color{grey} \scriptsize $b_1<225$}
 \psfrag{b1>225}{\color{grey} \scriptsize $b_1\gtrsim225$ (physical region)}
 \psfrag{validity}{\footnotesize validity of~\eqref{Equation:bcond}}
 \includegraphics[scale=0.56,clip]{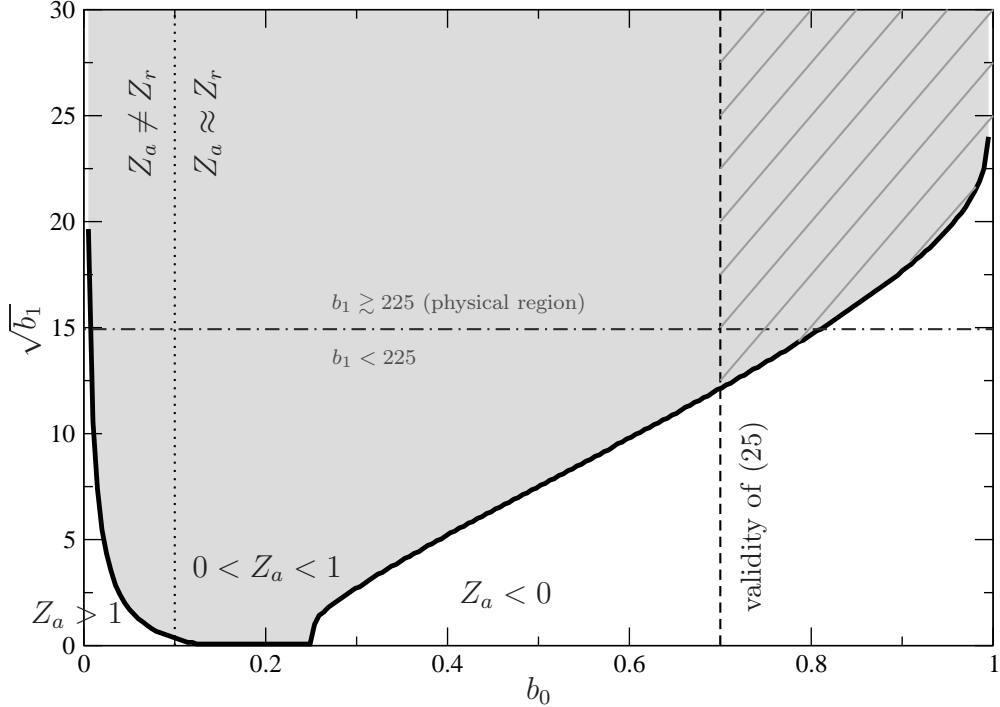}
 \caption{Consistency conditions for the Weinberg method in the $b_0$--$\sqrt{b_1}$--plane. 
The shaded grey area indicates $0<Z_a<1$, while the area to the right of the dotted horizontal line satisfies 
$Z_a\approx Z_r$. The area to the right of the dashed horizontal line satisfies the approximate condition~\eqref{Equation:bcond} for 
$b_0$. In the overlap of these regions indicated by the dashed area, the Weinberg formalism 
is applicable. This area also lies within the physical region of interest for the $Y(4660)$ ($b_1\gtrsim225$).}
\label{Graphic:Z-consistency}
\end{figure}

Another feature shown in Fig.~\eqref{Graphic:Z} is the fact that for certain parameters $Z_a$ exceeds its allowed 
interval $[0,1]$. This leads us back to question at which scales the initial approximation~\eqref{Equation:approx} 
is valid. The approximate condition~\eqref{Equation:bcond} for $b_0$ and the exact condition for $b_0$ and $b_1$
determined using Eq.~\eqref{Equation:consistency} as a consistency--check for the theory, are summarized in 
Fig.~\ref{Graphic:Z-consistency}. We show the consistency conditions for the Weinberg method in the 
$b_0$--$\sqrt{b_1}$--plane.\footnote{The dimensionless quantity $b_1$ has been replaced by its square root
to magnify the most relevant parameter region.}
The shaded grey area indicates the allowed region with $0<Z_a<1$, while the area to 
the right of the dotted horizontal line satisfies the consistency condition $Z_a\approx Z_r$. The area to the right of the 
dashed horizontal line satisfies the approximate condition \eqref{Equation:bcond}: $1\gg 1-b_0$.
Within the upper right corner (dashed area) both our approximate and numerically exact calculations 
indicate the applicability of the Weinberg method. Outside of this area it leads to inconsistent results and
can not be applied.
For the case of the $Y(4660)$, the Weinberg method is applicable in a significant part of the physical region 
of interest.  
This is due to the fact that the experimental/theoretical constraints $B_Y\lesssim20$~MeV and 
$|a_{\psi'K}|\lesssim0.1$~fm restrict the parameter $b_1=1/(2\mu_{\psi'K}\ a_{\psi'K}^2 \ B_Y)$ to be greater than 
approximately $225$. 

{\bf 5. Summary and Outlook:} 
In this work, we have investigated whether it is possible to distinguish between two- and three-body molecules 
and which scales govern this distinction. We identified the ratios of the characteristic energies of the two-body 
subsystems to the characteristic three-body energy $b_0$ and $b_1$ as the relevant scales. Moreover, we found
a parameter region where the two-body picture is appropriate and other regions where it leads to inconsistencies.
In particular, we have applied our formalism to the example of the $Y(4660)$. 
There, we found that a two-body picture is applicable in a significant part of the relevant region in the parameter space. 

We also note that our work can not yet be used to 
compare to experimental line shapes, especially since we assumed 
the $f_0(980)$ as a stable particle. In reality, due the decay $f_0(980)\to
\pi\pi$ the $Y(4660)$ becomes observable in $\psi'\pi\pi$ invariant mass
distributions and the resulting line shapes are believed to contain important
information on the nature of the $Y$~\cite{Guo:2008zg}.  Within the formalism
presented, this channel could be included via a complex scattering length of
the $\bar KK$ system --- the impact of unstable constituents
on the line shapes of particles is, e.g., discussed in Ref.~\cite{lineshapes}.
In addition, the $Y(4660)$ may also decay into other
channels. Currently there is a discussion, if the signal seen in
$\Lambda_c\bar \Lambda_c$, baptized $X(4630)$, has its origin in the
$Y(4660)$~\cite{LL1,LL2,LL3} --- this channel could be parametrized by an imaginary part
of the three--body interaction.

However, we do not expect that the possible extensions mentioned will
distort significantly the conclusions on the applicability of the Weinberg method.
Our analytical analysis summarized in Eq.~\eqref{Equation:bcond} shows that the binding 
energy of the $Y(4660)$ relative to the three-body threshold has to be
large compared to the molecular binding energy relative to the $\psi' f_0$ threshold,
$1\gg 1-b_0$. Moreover, we found that if the
scattering length in the other subsystems is large, this kind of analysis can
not be used and should either be replaced by a more complex, coupled channel
analysis or abandoned all together. However,  in a significant part of the
parameter space allowed for the $f_0(980)$ and the $Y(4660)$ 
shown in Fig.~\ref{Graphic:Z-consistency}, the Weinberg
method can be used to quantify the two--body molecular component. 
Because of the sizeable errors in the masses of the $Y(4660)$ and the $f_0(980)$ and the unknown 
$\psi'K$ scattering length, no definite conclusion about the nature of the $Y(4660)$ can be reached at present.

Various extensions of our approach are possible.
In principle, we could also integrate spin--effects, higher derivatives in the fields and, via photon--coupling, even 
charge--dependent interactions into the Lagrangian~\eqref{Equation:Lagrangian}. Clearly all these extensions imply 
additional parameters in our theory which would have to be determined.
Furthermore, we could scan the field of possible hadronic molecules for dimer-- and three--particle--candidates. 
An interesting application is the $X(3872)$--meson as a $DD\pi$ system. 
Since the $D\pi$--dimer 
could only appear in $p$-wave scattering, higher derivatives in the fields would have to be included.

\noindent 
{\bf Acknowledgments}

This work was supported by the DFG through SFB/TR 16 \lq\lq Subnuclear structure of matter'', 
by the Helmholtz Association through funds provided 
to the virtual institute ``Spin and strong QCD'' (VH-VI-231),
and by the BMBF under contract No. 06BN9006 \lq\lq Strong interaction studies for FAIR''.

\end{document}